\documentclass[12pt]{article}
\usepackage{fancyhdr}
\usepackage[margin=1in,nohead]{geometry}
\usepackage{wrapfig}
\usepackage{amssymb}
\usepackage[flushleft]{threeparttable}
\usepackage{graphics,graphicx}
\usepackage{arydshln}

{\begin{list}{}%
         {\setlength{\leftmargin}{#1}}%
         \item[]%
}
{\end{list}}

\newcommand{\sol}{$_{\odot}$}

\begin{document}
\centerline{\bf The Case for a James Webb Space Telescope
  Extragalactic Key Project}
  \vspace{0.5mm}
\centerline{\it Steven L.\ Finkelstein$^{1}$, James Dunlop$^{2}$, Olivier Le F\`evre$^{3}$ \& Stephen Wilkins$^{4}$}

\centerline{\it \footnotesize $^{1}$Department of Astronomy, The University of Texas at Austin, USA}
\centerline{\it \footnotesize $^{2}$Institute for Astronomy, Royal Observatory, University of Edinburgh, UK}
\centerline{\it \footnotesize $^{3}$Laboratoire d'Astrophysique de Marseille, Aix-Marseille Universit\'{e}, France}
\centerline{\it \footnotesize $^{4}$Astronomy Centre, Department of Physics and Astronomy, University of Sussex, UK}
\vspace{2mm}

\emph{This document summarizes and develops the results of a discussion session at the "Exploring the Universe with JWST" meeting at ESA-ESTEC in October 2015.}
\vspace{4mm}

The upcoming launch of the {\it James Webb Space Telescope} ({\it JWST}) in less
than three years is certain to bring a revolution in our understanding
of galaxy evolution.  As the first proposals will be due in a
little over two years, the time is ripe to take a holistic look at the
science goals which the community would wish to accomplish with this
observatory.  Contrary to our experiences with the {\it Hubble Space
Telescope} ({\it HST}), which has now operated successfully for over two decades due to several timely servicing missions, the lifetime of {\it JWST} is finite and
relatively short with a lifetime requirement of five years, and a ten-year goal.  In this document we highlight the (non-local)
extragalactic science goals for {\it JWST} and describe how a concerted community effort will best address these, ensuring that the desired survey can be completed during the {\it JWST} mission.

One of the key extragalactic science goals for {\it JWST} is to discover the first
galaxies to exist in the distant universe.  Wavelength and aperture
limitations restrict {\it HST} studies to $z < 10$, yet the relatively bright and massive galaxies we can now find out to such distances hint at a tantalizing plethora of sources to discover in the even more distant past.  Deep imaging with {\it JWST} NIRCam out to $\lambda \simeq 3–-5\,\mu$m will allow the discovery of star-forming galaxies out to possibly $z\sim 20$ (and certainly to $z\sim15$).  In addition to probing the physics of galaxy formation only $\sim$200 Myr after the Big
Bang, such studies will allow us to trace the evolution of the cosmic
star-formation rate density to the earliest cosmic times, search for
evidence of the first stars (via enrichment signatures, or possibly
supernovae), and provide extremely tight constraints on the
contribution of galaxies to reionization (both through tracking the evolving galaxy population back to earlier times, and also by probing deeper at the currently studied epoch
corresponding to redshifts 6 $< z <$ 10).  These topics are at the forefront of current
astrophysical research, being prominently featured in the recent US
decadal survey as well as NASA’s Cosmic Origins goals.  Most crucially, "The End of the Dark Ages: First Light and Reionization" and "The Assembly of Galaxies" are the first two of the four primary {\it JWST} Science Goals\footnote[1]{http://www.stsci.edu/jwst/doc-archive/science-requirements.pdf}.

Another key goal for {\it JWST} is to perform a comprehensive cosmic census of
galaxies.  A deep 1—-10 $\mu$m imaging survey would allow the selection of
galaxies via stellar mass, rather than star-formation activity.  By probing
stellar emission over rest-frame near-infrared wavelengths,
such a survey would be sensitive to all galaxies down to a specific
stellar mass limit, regardless of the current level of star formation.
This is in stark contrast to essentially all current studies at $z > 6$,
which rely on rest-frame ultraviolet observations to select galaxies,
and thus are not sensitive to galaxies more than 100 Myr after their most recent
episode of star formation.   Such a survey would allow detailed
investigations into the evolution of the total stellar mass density,
the star-formation duty cycle, and a robust search into the
progenitors of today's massive galaxies at early times.

The science goals discussed above, which cover some of the key
unanswered questions in galaxy evolution, can all be addressed with
the same survey: a deep 1—-10 $\mu$m imaging survey.  Such a survey could be utilized by any science investigation at $0.5 < z < 15$ that is not reliant on a specific region of the sky.  This survey can also be optimized for supernovae (SNe) searches, allowing the discovery of SNe to $z >$ 2, and mitigating the effect of dust on $z <$ 2 SNe light curves.
We now turn our attention to investigating the
parameters for this potential survey.  As we note below, such a survey requires both NIRCam and MIRI, and thus stands to benefit substantially by the ability to observe with both instruments in parallel.  Likewise, performing some of the required imaging in parallel with prime NIRSpec surveys can also increase the efficiency of the proposed survey.
Clearly spectroscopy would also be a desired component.  However, this may be more difficult to implement in a key project simultaneously with imaging due to the necessity of target selection.  However, one path to integrate spectroscopy with imaging in an efficient manner would be to include a grism component (with NIRSpec follow-up spectroscopy of interesting sources being pursued by the community), at the cost of increasing the program size.

One possible design would be to survey the CANDELS\footnote[2]{The Cosmic Assembly Near-infrared Deep Extragalactic Legacy Survey; candels.ucolick.org}
fields with a wedding-cake strategy.  The choice of several fields distributed
around the sky mitigates cosmic variance and eases the
scheduling of such a program.  In the era of {\it JWST} these particular fields may be less unique (though the in-place X-ray and infrared imaging will be useful) and other fields may be easier to schedule, thus other field options can certainly be considered.  We base our straw-man survey design
(which we stress is simply illustrative of the possible total time
needed, and does not represent a thorough investigation into specific
science requirements) on achieving the above science goals.  The deep survey must be sensitive enough to detect rest-frame UV emission
from a M$_{\ast}$$\sim$2$\times$10$^7$ M\sol\ star-forming galaxy at $z\sim$14
(m$_\mathrm{UV} \sim$ 30.5), and rest-frame optical/near-infrared emission from a M$_{\ast}$$\sim$2$\times$10$^9$
M\sol\ post-starburst galaxy at $z=$7 (m$_\mathrm{NIR} \sim$ 27).  The wide
portion must still be sensitive enough to detect sub-L$^{\ast}$ galaxies at
$z>$10 with NIRCam (m$\sim$29) and sub-L$^{\ast}$ galaxies at $z\sim$6 with MIRI (m$\sim$26).
Investigating several methods of tiling, we find that we can cover the
majority of one CANDELS field with the configuration shown in Figure
1.  This consists of a NIRCam survey with one deep pointing (m=30.5,
shown in red), four medium-deep pointings (m=30, yellow), and eight
wide pointings (m=29, green), and a MIRI survey with three deep
pointings (m=27), 12 medium-deep pointings (m=26.5), and 28 wide
pointings (m=26).  

\begin{figure}[ht]
\centering
\includegraphics[width=4.0in]{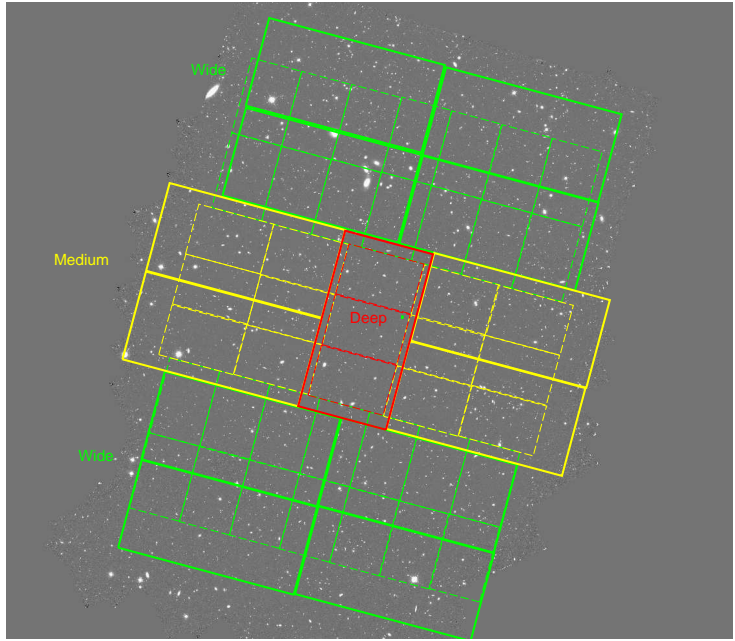}
\vspace{-5mm}
\caption{Straw-man survey design for coverage of the GOODS-S field.
The large solid rectangles represent the NIRCam
2.2$^{\prime}\times4.4^{\prime}$ field-of-view, while the dashed
smaller rectangles denote the MIRI 1.25$^{\prime}\times1.88^{\prime}$
field-of-view.  The red, yellow and green colors denote the deep,
medium and wide components, respectively.  The total area covered by
NIRCam is 125 arcmin$^2$, and by MIRI is 101.5 arcmin$^2$.}
\end{figure}

We use the online {\it JWST} prototype exposure time calculator\footnote[3]{http://jwstetc.stsci.edu/etc/} to estimate the exposure times for
this straw-man survey.  To reach 5$\sigma$ limiting magnitudes of
30.5, 30 and 29 in seven NIRCam filters (Table 1) requires a
total exposure time of 88, 35 and 6 hours, respectively (obtaining one
short and one long channel filter simultaneously).  To reach the same
significance threshold with MIRI in the F560W and F770W filters at
m=27, 26.5, and 26 requires 18, 7.5 and 3 hours, respectively.
Combining these exposure times with the number of pointings in the
previous paragraph, we find total exposure times of 276 hours of
integration with NIRCam and 160 hours with MIRI (assuming a
conservative 30\% savings on MIRI exposure time via coordinated
parallels; the full exposure time needed is 228 hours).  Therefore the total
cost to survey one CANDELS field would be 436 hours.  To perform such
a survey over five such fields would therefore be 2180 hours (exclusive of overheads).  

Although we only present one possible plan here, any survey which
wishes to answer all of the outlined key science goals will likely require 
over 1000 hours.  A 1000-2000 hour survey is clearly an enormous
investment of telescope time, representing 20-40\% of the available
hours in one year, and 4-8\% of the total number of hours over a five
year mission duration (assuming 5000 hr/yr).  
A coordinated community survey is the best path forward to ensure this program maximizes a combination of science goals, and hence maximizes the science return for the {\it JWST} observatory. Large imaging surveys on {\it HST} (e.g., GOODS, CANDELS, COSMOS, the Hubble Frontier Fields) are supported by a large community which releases and exploits the data in a timely manner. Competition in doing similarly large surveys with {\it JWST} makes little sense as the parameter space is well-defined already and will be common to any proposing team (i.e., a wedding cake approach).  If no such initiative is in place, then it is likely that such a program will be doomed to the benefit of smaller, less ambitious programs that will affect the overall science return for a broad community.

While such a survey is clearly essential for maximizing our use of
{\it JWST}, it may not be prudent to set off on this path immediately.  As
a new observatory, there may be many aspects of how the telescope
operates which are currently unknown, yet may influence the design of
such a survey (including uncertainties on the expected sensitivities,
whether the cameras integrate down as expected, etc.).  These unknowns
will be explored during the first cycle, by commissioning,
guaranteed time (GTO) and early release science (ERS) observations.  Therefore,
such a survey as we propose here may be more timely to begin in Cycle
2, when the Cycle 1 programs can provide direct input on the planning
of this large endeavor.  Additionally, GTO, ERS, and Cycle 1 programs 
will likely accomplish some of the required imaging, 
providing testbeds and seeds for larger surveys.

We propose two options.  The first would be for the Space
Telescope Science Institute to perform this survey as a “key”
project, with the project time coming off the top of the entire
mission time budget.  This could be done in an institute-led,
community-driven fashion, similar to the Hubble Frontier Fields (HFFs),
with the construction of an advisory committee which seeks input from
the community on the science goals and survey design.  In contrast to
the HFFs, a survey of such importance should have several levels of
iteration and engagement with the community before a design is
finalized.  We prefer this option as it would result in a public
survey maintaining the spirit of openness established through the existing legacy fields.
Should such a program occur, we stress it is of the utmost importance
to ensure that adequate funding is available through channels such as the archival program in the US, and national or EU (e.g. ESA or European Research Council) funding in Europe.
It would also be advisable to organize science working groups to assist with the training and
networking for junior scientists, which would occur naturally
should this survey occur in a more traditional PI-led fashion.  A
second option would be for STScI to have a proposal call similar to
{\it HST}’s highly successful Multi-Cycle Treasury program.  If this
call had adequate resources (several thousand hours to allocate over
several years), it would provide a mechanism for a comprehensive galaxy survey
to succeed in a competitive environment.
\vspace{1mm}

We thank the attendees of what proved to be a productive discussion session at the ESTEC {\it JWST}
meeting, and acknowledge useful discussions with several colleagues over the past
weeks which have produced further useful input and feedback.

\begin{table}[!h]
\centering
\begin{threeparttable}
\caption{\bf Strawman Survey Design}
\begin{tabular*}{0.8\textwidth}
   {@{\extracolsep{\fill}}cccc}
\hline
\hline
\multicolumn{1}{c}{Filter} & \multicolumn{1}{c}{t($m_{5\sigma} =$ 30.5)} & \multicolumn{1}{c}{t($m_{5\sigma} =$ 30)} & \multicolumn{1}{c}{t($m_{5\sigma} =$ 29)}\\
\multicolumn{1}{c}{$ $} & \multicolumn{1}{c}{(ksec)} &
\multicolumn{1}{c}{(ksec)} & \multicolumn{1}{c}{(ksec)}\\
\hline
NIRCam  F090W&105&42&7\\
NIRCam  F115W&85&34&6\\
NIRCam  F150W&70&28&4.5\\
NIRCam  F200W&50&20&3.5\\
\hdashline%[0.5pt/5pt]
NIRCam  F277W&50&20&3.5\\
NIRCam  F356W&70&30&5\\
NIRCam  F444W&200&80&13\\
\hline
Total NIRCam (hr)&88&35&6\\
\hline
\hline
\multicolumn{1}{c}{Filter} & \multicolumn{1}{c}{t($m_{5\sigma} =$ 27)} & \multicolumn{1}{c}{t($m_{5\sigma} =$ 26.5)} & \multicolumn{1}{c}{t($m_{5\sigma} =$ 26)}\\
MIRI F560W&16&6.5&2.5\\
MIRI F770W&50&21&8\\
\hline
Total MIRI (hr)&18&7.5&3\\
\hline
\hline
\end{tabular*}
\begin{tablenotes}
  \small
\item The total NIRCam time denotes the total
  in each NIRCam channel (short and long, split by the dashed line), which is approximately equal for this filter set.
\end{tablenotes}
\end{threeparttable}
\end{table}

\end{document}